\documentclass[final]{aipproc}
\usepackage{graphicx,amsmath,amsfonts,amssymb,mathrsfs}

\layoutstyle{8x11double}

\newcommand{\ie}{{\it i.e.}}

\newcommand{\eg}{{\it e.g.}}
\newcommand{\cf}{{\it cf.}}

\newcommand{\etc}{{\it etc.}}
\newcommand{\eq}{Eq.}

\newcommand{\fig}{Figure}

\newcommand{\Ref}{Ref.}
\newcommand{\Refs}{Refs.}

\newcommand{\Tab}{Table}

\newcommand{\stheta}{\sin^22\theta_{13}}
\newcommand{\deltacp}{\delta_{\mathrm{CP}}}

\newcommand{\equ}[1]{\eq~(\ref{equ:#1})}
\newcommand{\figu}[1]{\fig~\ref{fig:#1}}

\newcommand{\bi}{\begin{itemize}}
\newcommand{\ei}{\end{itemize}}

\begin{document}

\title{Systematic Model Building Based on \\ Quark-Lepton Complementarity Assumptions}

\classification{14.60.Pq
%<Replace this text with PACS numbers; choose from this list:
%                \texttt{http://www.aip..org/pacs/index.html}>
}

\author{Walter Winter}{
  address={Institut f{\"u}r theoretische Physik und Astrophysik, Universit{\"a}t W{\"u}rzburg \\
Am Hubland, D-97074 W{\"u}rzburg, Germany}
}

\begin{abstract}
In this talk, we present a procedure to systematically generate a large number of
valid mass matrix textures from very generic assumptions. Compared to plain anarchy arguments, we postulate some structure for the theory, such as a possible connection between quarks and leptons, and a mechanism to generate  flavor structure. We illustrate
how this parameter space can be used to test the exclusion power of future experiments,
and we point out that one can systematically generate embeddings in $Z_N$ product flavor symmetry 
groups. 
\end{abstract}

\maketitle

%%%%%%%%%%%%%%%%%%%%%%%%%%%%%%%%%%%%%%%%%%%%
%% MAINMATTER
%%%%%%%%%%%%%%%%%%%%%%%%%%%%%%%%%%%%%%%%%%%%

In the literature, many approaches to a theory of lepton masses and mixings have been studied,
such as descriptions by textures, GUTs, flavor symmetries, \etc\ (see \Ref~\cite{Albright:2006cw}
for a dedicated study). Most of these approaches are presented as possible, consistent theories predicting observables passing the current experimental constraints. However, it is difficult to obtain information on the parameter space as a whole which experiments are going to test, 
because it is not possible to compare individual models on a 
statistical basis. One possibility for such a test is generating a large sample
of possibilities from very generic arguments. The most generic assumption, one can probably
think of, are anarchic entries for the Yukawa couplings~\cite{Hall:1999sn%,Haba:2000be,deGouvea:2003xe
}.
This procedure leads to very generic (and model-independent) distributions of the observables.
However, one has to give up the belief in an underlying structure, such as a flavor symmetry. In this talk, we discuss a generic possibility to create a large
parameter space of (valid) models using somewhat more structure. Our generic assumptions will be motivated
by a possible connection between quarks and leptons, such as (see, \eg, \Refs~\cite{Rodejohann:2003sc%,Smirnov:2004ju,Minakata:2004xt,Raidal:2004iw,Datta:2005ci,Li:2005ir,Xing:2005ur,Everett:2005ku
})
\begin{equation}
\theta_{12} + \theta_C \simeq \pi/4 \simeq \theta_{23} \, .
\label{equ:qlc}
\end{equation}
This formula is often referred to as ``quark-lepton complementarity'' (QLC).
In addition, we will illustrate how this approach can be systematically linked to flavor symmetries.

In our bottom-up approach~\cite{Plentinger:2006nb,Plentinger:2007px,Winter:2007yi}, we start with our generic assumptions in order to generate neutrino mass matrices including order one coefficients. We compute the neutrino oscillation observables from these matrices and check for compatibility with data. We call these matrices {\em realizations}. For all valid realizations, we then identify the leading order structure, which depends on the generic assumptions. We call these structures {\em textures}, since they contain the information on the origin of the Yukawa couplings. For example, for random mass matrices, one could identify texture zeros as the structure in the mass matrices. These textures can then be embedded in {\em models} in a specific framework. For example, one could use flavor symmetries to produce the structure in the textures. Note that there is a $1:n$ correspondence between realizations and textures, and there can be a $m:1$ correspondence between models and one specific texture. At the end, the interpretation of our results can be done in the reverse direction: From a model one obtains the texture, and for a (valid) texture, one immediately knows that there is a set of valid order one coefficients to reproduce this structure. There are several advantages of our approach. First, depending on the specific mechanism, one may not need diagonalization of the effective neutrino mass matrix, which simplifies the whole process tremendously. Second, from the generic assumptions, it is difficult to predict the outcome. Therefore, it is also difficult to introduce bias at this stage.
Third, we will construct {\em all} possibilities given a set of generic assumptions, which will produce all new possibilities (textures, models) given our set of assumptions. And fourth, it is, of course, the objective of any systematic approach to study the parameter space. As a spin-off, one can easily demonstrate  how experiments can test this parameter space most efficiently.

\begin{figure}[t!]
\hspace*{0.3cm}
$
M_\nu^{\rm Maj} = \left(
\begin{array}{ccc}
\epsilon^2 & \epsilon^2 & 0 \\
\epsilon^2 & \epsilon & \epsilon^2 \\
0 & \epsilon^2 & 1 
\end{array}
\right)
$
 \raisebox{-1.5cm}{\includegraphics[width=13cm]{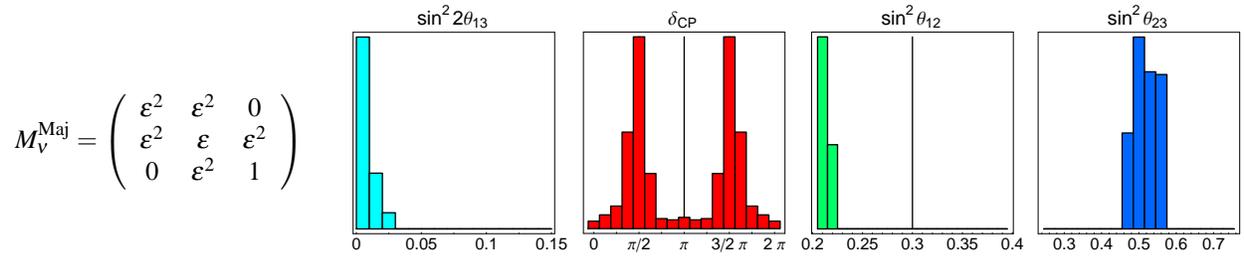}} \\
\caption{\label{fig:predhist}An example for a neutrino mass texture, and the corresponding
distributions of the observables from the realizations leading to this texture. For details and
more examples, see \Ref~\cite{Winter:2007yi}.}
\end{figure}

For the following, we will choose a specific generic assumption, which we have called ``extended quark-lepton complementarity''. Based upon \equ{qlc}, we postulate that all masses and mixings in the quark and lepton sectors be described by powers of $\epsilon \simeq \theta_C$ as a potential remnant of a quark-lepton unified theory, \ie, by $\epsilon^0$, $\epsilon^1$, $\epsilon^2$, \etc , where $\epsilon^0$ corresponds to maximal mixing for the mixing angles. One can easily demonstrate that all quark and lepton mass hierarchies can be described by powers of $\epsilon$, such as $\epsilon^2:\epsilon:1$ for a normal neutrino mass hierarchy, and all mixings as well~\cite{Plentinger:2007px}. For example, the Wolfenstein parameter $\lambda\simeq \epsilon$ allows for a parameterization of $V_{\mathrm{CKM}}$ (without phases), and the lepton mixing might be described by $U_{\mathrm{PMNS}} = V_{\mathrm{CKM}}^\dagger V_{\mathrm{Bimax}}$ as a product of $V_{\mathrm{CKM}}$ and a mixing matrix with two maximal mixing angles as a special case~\cite{Jezabek:1999ta%,Giunti:2002pp,Frampton:2004ud
}. Using our generic assumptions, we generate all possible realizations, filter the realizations compatible with data, and calculate the textures by identifying the leading order in $\epsilon$. Note that, compared to the texture zero case, we allow for more structure in the textures: The entries can be $1$, $\epsilon$, $\epsilon^2$, or $0$, where $0$ corresponds to $\mathcal{O}(\epsilon^3)$.\footnote{Because of the current experimental precision, it does not make sense to go to higher orders in $\epsilon$ yet.} Of course, keeping this type of information in the textures is only relevant as long as one wants to go one step further: We will illustrate how the powers of $\epsilon$ can be connected to specific models, such as discrete flavor symmetries.

Let us now introduce our procedure for the (effective) $3\times3$ case.
First, we generate all possible pairs $\{ U_\ell$, $U_\nu \}$ 
using the standard parameterization for the unitary mixing matrices.
We  generate all possible mixing angles $\sin \theta_{ij}^\alpha \in \{ 1/\sqrt{2}, \epsilon, \epsilon^2, 0 \}$, and all possible real phases ($0$ or $\pi$). 
Then we calculate $U_{\rm PMNS}$ by 
\begin{equation}
U_{\rm PMNS}=U_\ell^\dagger U_\nu \, ,
\label{equ:upmns}
\end{equation}
read off the mixing angles and
observable phases, and select those
realizations with mixing angles being compatible with current data at 
the $3 \sigma$ confidence level (\cf, \Ref~\cite{Plentinger:2006nb}).
For each valid realization, we then find, for instance, the corresponding 
Majorana mass matrix {\em texture} by 
computing 
\begin{equation}
M_\nu^{\rm Maj} = U_\nu M_\nu^{\rm diag} U_\nu^T \, ,
\label{equ:maj}
\end{equation}
expanding in $\epsilon$, and by identifying the first non-vanishing coefficient.
Here the assumptions for $M_\nu^{\rm diag}$ are taken from extended QLC as well,
such as $m_1:m_2:m_3=\epsilon^2:\epsilon:1$ for the normal mass hierarchy.
In this procedure, diagonalization is not necessary compared to generating
the entries in $M_\nu^{\rm Maj}$ directly.

As a result, a systematically generated set of textures and sum rules is
obtained~\cite{Plentinger:2006nb}.  For example,  one obtains a 
``diamond-shaped'' texture which
 can be produced by the two maximal mixing angles
in the lepton sector ($\theta_{12}^\ell$ and $\theta_{23}^\ell$)
and one maximal mixing angle in the neutrino sector ($\theta_{13}^\nu$).
In addition, one can study the distributions of observables obtained
from the set of realizations. For example, a large mixing $\stheta$
close to the current bound is indeed preferred from the complete
set of valid realizations. Furthermore, since
$\theta_{12}$ can only be obtained by the matrix multiplication
in \equ{upmns} from maximal and Cabibbo-like mixing angles,
future precision measurements of $\theta_{12}$ will exert pressure 
on the parameter space. 

As the next step, one can introduce complex phases in the effective $3\times3$
mechanism by systematically generating all possible phases with
uniform distributions~\cite{Winter:2007yi}. If one then
studies the distributions of observables as a function of the texture, one finds
that these distributions are very texture dependent. For example, \figu{predhist}
shows a texture preferring small $\stheta$, maximal CP violation, $\theta_{23}$
close to the best-fit value, and $\theta_{12}$ close to the currently allowed lower
bound. For the present example, a detection of large $\stheta$, or a confirmation of $\theta_{12}$ 
close to the current best-fit value would clearly disfavor the texture.
It turns out that there are many such qualitatively different cases
as a function of the texture one uses. In addition, one finds peaks
 close to CP conservation in some of the $\deltacp$ distributions, where the
deviations from $0$ or $\pi$ are of the order $\theta_C$. Such peaks may motivate a $\deltacp$ precision
of future high precision measurements at the level of $\theta_C\simeq11^\circ$, as it could
be obtained at a neutrino factory ($1\sigma$)~\cite{Huber:2004gg}.

\begin{table}[t]
\begin{tabular}{ccc}
\hline
$M_\ell$ & $M_D$ & $M_R$ \\
\hline
$\left(
\begin{array}{ccc}
0 & \epsilon^2 & 1 \\
0 & \epsilon^2 & \epsilon \\
0 & \epsilon^2 & 1 \\
\end{array}
\right)$ &
$\left(
\begin{array}{ccc}
\epsilon & 0 & 0 \\
\epsilon & 1 & \epsilon \\
\epsilon & 1 & 0 \\
\end{array}
\right)$ &
$\left(
\begin{array}{ccc}
\epsilon & \epsilon & 0 \\
\epsilon & 1 & 0 \\
0 & 0 & 1 \\
\end{array}
\right)$ \\
\hline
\multicolumn{3}{p{5cm}}{
\vspace*{-0.7cm} 
\begin{eqnarray}
(\nu_1^c,\nu_2^c,\nu_3^c) & = & (1,0,1) \quad (0,3,2) \quad (3,3,0) \nonumber \\
(\ell_1,\ell_1,\ell_3) & = & (4,3,2) \quad (0,1,0) \quad (0,2,2) \nonumber \\
(e_1^c,e_2^c,e_3^c) & = & (3,0,2) \quad (2,0,2) \quad (1,2,0) \nonumber 
\end{eqnarray}
\vspace*{-0.7cm}
} \\
\hline
\end{tabular}
\caption{\label{tab:st} First row: A possible texture obtained in the seesaw case (for small $\stheta$ and a normal hierarchy)~\cite{Plentinger:2007px}. Second row: A set of possible quantum numbers
for a $Z_5 \times Z_4 \times Z_3$ flavor symmetry. Note that common pre-factors can be absorbed in the absolute mass scale, and that a texture zero corresponds to $\mathcal{O}(\epsilon^3)$.
} 
\end{table}

Our procedure can be extended to the seesaw mechanism by parameterizing the $6\times 6$ 
neutrino mass matrix in terms of three unitary matrices and the corresponding mass eigenvalues in a similar way~\cite{Plentinger:2007px}. Similar to the $3\times 3$ case, the charged lepton sector is
 {\em not} assumed to be diagonal. We find $1 \, 981$ different textures generated with the
assumption of $\stheta$ small and of a normal neutrino mass hierarchy. One example, which can be realized with $(\theta_{12},\theta_{13},\theta_{23})=(33^\circ,0.2^\circ,52^\circ)$, is shown in \Tab~\ref{tab:st}
(upper row). Special cases, such as
$M_D$ symmetric, are rare in our sample. However, about a quarter of all realizations lead to
$M_R$ being almost diagonal. For the mass hierarchies, we obtain many cases with mild hierarchies
in $M_D$ and $M_R$. Not surprisingly, charged lepton mixing is quite substantial in many cases. In fact, for about one third of all possible realizations, there are three maximal mixing angles in $U_\ell$.

Let us now discuss what the Yukawa coupling structure, such as the one shown in \Tab~\ref{tab:st}, is actually good for.
For example, masses for quarks and leptons may arise from higher-dimension terms via
the Froggatt-Nielsen mechanism~\cite{Froggatt:1978nt} in combination with a flavor symmetry:
\begin{equation}
\mathcal{L}_\mathrm{eff} \sim \langle H \rangle \, \epsilon^n \, \bar{\Psi}_L \Psi_R \, .
\label{equ:fn}
\end{equation}
In this case, $\epsilon$ becomes meaningful in terms of a small
parameter $\epsilon=V/M_F$ which controls the flavor symmetry breaking.\footnote{
Here $V$ are universal VEVs of SM singlet scalar ``flavons'' that break the flavor symmetry,
and $M_F$ refers to the mass of superheavy fermions, which are charged under the
flavor symmetry. The SM fermions are given by the $\Psi$'s.} 
The integer power of $\epsilon$ is solely determined by the
quantum numbers of the fermions under the flavor symmetry (see, \eg, \Refs~\cite{Plentinger:2007px,Enkhbat:2005xb}). 
Using discrete Abelian $Z_N$ product flavor symmetry groups, one can now try to
reproduce our textures by calculating the specific embeddings~\cite{phaseprep}.
For instance, one can discuss how much complexity is actually needed to reproduce
almost all of our $1 \, 981$ textures. While one can find embeddings
for simple textures already for symmetries such as $Z_4$,
it turns out, that one can find model embeddings for about $60\%$ of our textures
for $Z_5 \times Z_4 \times Z_3 \times Z_2$. Therefore, with that amount of complexity,
one can almost reproduce any viable texture. As one non-trivial example,
consider the $Z_5 \times Z_4 \times Z_3$ discrete symmetry embedding for the texture
in \Tab~\ref{tab:st} (upper row): One possible set of quantum numbers is given in \Tab~\ref{tab:st}, lower row.

In summary, we have demonstrated how our automatic procedure can be used to 
automatically generate valid textures, corresponding order one coefficients
compatible with data, and specific models including the charge assignments
at the example of a discrete $Z_N$ product flavor symmetry groups.
While the generation is bottom-up, 
the interpretation of our results can be performed in a top-down fashion:
A model predicts a form of the mass matrix (texture), and this form is known to fit data with
proper order one coefficients (realization). As potential applications, our approach can be
used of parameter space studies for the observables or for theories (models), as well
as for finding new models.  As a spin-off, the exclusion power of experiments in
the generated parameter space can easily be tested.

{\bf Acknowledgments}
 I would like to acknowledge support from the Emmy Noether program of Deutsche Forschungsgemeinschaft.

%\bibliographystyle{aipproc}  
%\bibliography{references}

\vspace*{-1cm}

\begin{thebibliography}{22}
\expandafter\ifx\csname natexlab\endcsname\relax\def\natexlab#1{#1}\fi
\providecommand{\enquote}[1]{``#1''}
\expandafter\ifx\csname url\endcsname\relax
  \def\url#1{\texttt{#1}}\fi
\expandafter\ifx\csname urlprefix\endcsname\relax\def\urlprefix{URL }\fi
\providecommand{\eprint}[2][]{\url{#2}}

\bibitem[Albright and Chen(2006)]{Albright:2006cw}
C.~H. Albright, and M.-C. Chen  (2006), \eprint{hep-ph/0608137}.

\bibitem[Hall et~al.(2000)]{Hall:1999sn}
L.~J. Hall, H.~Murayama, and N.~Weiner, \emph{Phys. Rev. Lett.} \textbf{84},
  2572--2575 (2000), \eprint{hep-ph/9911341};
N.~Haba, and H.~Murayama, \emph{Phys. Rev.} \textbf{D63}, 053010 (2001),
  \eprint{hep-ph/0009174};
A.~de~Gouvea, and H.~Murayama, \emph{Phys. Lett.} \textbf{B573}, 94--100
  (2003), \eprint{hep-ph/0301050}.

\bibitem[Rodejohann(2004)]{Rodejohann:2003sc}
W.~Rodejohann, \emph{Phys. Rev.} \textbf{D69}, 033005 (2004),
  \eprint{hep-ph/0309249};
A.~Y. Smirnov  (2004), \eprint{hep-ph/0402264};
H.~Minakata, and A.~Y. Smirnov, \emph{Phys. Rev.} \textbf{D70}, 073009 (2004),
  \eprint{hep-ph/0405088};
M.~Raidal, \emph{Phys. Rev. Lett.} \textbf{93}, 161801 (2004),
  \eprint{hep-ph/0404046};
A.~Datta, L.~Everett, and P.~Ramond, \emph{Phys. Lett.} \textbf{B620}, 42--51
  (2005), \eprint{hep-ph/0503222};
N.~Li, and B.-Q. Ma, \emph{Phys. Rev.} \textbf{D71}, 097301 (2005),
  \eprint{hep-ph/0501226};
Z.-z. Xing, \emph{Phys. Lett.} \textbf{B618}, 141--149 (2005),
  \eprint{hep-ph/0503200};
L.~L. Everett, \emph{Phys. Rev.} \textbf{D73}, 013011 (2006),
  \eprint{hep-ph/0510256}.

\bibitem[Plentinger et~al.(to appear{\natexlab{a}})]{Plentinger:2006nb}
F.~Plentinger, G.~Seidl, and W.~Winter, \emph{Nucl. Phys.} \textbf{B} (to
  appear{\natexlab{a}}), \eprint{hep-ph/0612169}.

\bibitem[Plentinger et~al.(to appear{\natexlab{b}})]{Plentinger:2007px}
F.~Plentinger, G.~Seidl, and W.~Winter, \emph{Phys. Rev.} \textbf{D} (to
  appear{\natexlab{b}}), \eprint{arXiv:0707.2379 [hep-ph]}.

\bibitem[Winter(2007)]{Winter:2007yi}
W.~Winter  (2007), \eprint{arXiv:0709.2163 [hep-ph]}.

\bibitem[Jezabek and Sumino(1999)]{Jezabek:1999ta}
M.~Jezabek, and Y.~Sumino, \emph{Phys. Lett.} \textbf{B457}, 139--146 (1999),
  \eprint{hep-ph/9904382};
C.~Giunti, and M.~Tanimoto, \emph{Phys. Rev.} \textbf{D66}, 113006 (2002),
  \eprint{hep-ph/0209169};
P.~H. Frampton, S.~T. Petcov, and W.~Rodejohann, \emph{Nucl. Phys.}
  \textbf{B687}, 31--54 (2004), \eprint{hep-ph/0401206}.

\bibitem[Huber et~al.(2005)]{Huber:2004gg}
P.~Huber, M.~Lindner, and W.~Winter, \emph{JHEP} \textbf{05}, 020 (2005),
  \eprint{hep-ph/0412199}.

\bibitem[Froggatt and Nielsen(1979)]{Froggatt:1978nt}
C.~D. Froggatt, and H.~B. Nielsen, \emph{Nucl. Phys.} \textbf{B147}, 277
  (1979).

\bibitem[Enkhbat and Seidl(2005)]{Enkhbat:2005xb}
T.~Enkhbat, and G.~Seidl, \emph{Nucl. Phys.} \textbf{B730}, 223--238 (2005),
  \eprint{hep-ph/0504104}.

\bibitem[Plentinger et~al.(in preparation)]{phaseprep}
F.~Plentinger, G.~Seidl, and W.~Winter  (in preparation).

\end{thebibliography}

\renewcommand\refname{}

\end{document}